\documentclass[twocolumn,showpacs,preprintnumbers,prb]{revtex4-1}
\usepackage{amssymb}
\usepackage[dvips]{color}
\usepackage{epsfig}
\usepackage{dcolumn}
\usepackage{bm}
\usepackage[normalem]{ulem}

\def\r{\rho}
\def\w{\omega}

\def\s{\sigma}
\def\S{\Sigma}

\begin{document}

\title{Scanning Tunneling Microscope Operating as a Spin-diode}

\author{P. H. Penteado$^{1,2}$, F. M. Souza$^2$, A. C. Seridonio$^3$, E. Vernek$^2$, and J. C. Egues$^1$}
\affiliation{$^1$Instituto de F\'isica de S\~ao Carlos, Universidade de S\~ao Paulo, 13560-970, S\~ao Carlos, SP, Brazil\\$^2$Instituto de F\'isica, Universidade Federal de Uberl\^andia, 38400-902, Uberl\^{a}ndia, MG, Brazil\\$^3$Departamento de F\'isica e Qu\'imica, Universidade Estadual Paulista J\'ulio de Mesquita Filho, 15385-000, Ilha Solteira, SP, Brazil.} 

\begin{abstract}
We theoretically investigate spin-polarized transport in a system composed of a ferromagnetic Scanning Tunneling Microscope (STM) tip coupled
to an adsorbed atom (adatom) on a host surface. Electrons can tunnel directly from the tip to the surface or 
via the adatom. Since the tip is ferromagnetic and the host surface (metal or semiconductor) is non-magnetic we obtain a spin-diode effect when the adatom is in the regime of single occupancy. This effect leads to an unpolarized current for direct bias ($V>0$) and polarized current for reverse ($V<0$) bias voltages, if the tip is nearby the adatom. Within the nonequilibrium Keldysh technique we analyze the interplay between the lateral displacement of the tip and the intra adatom Coulomb interaction on the spin-diode effect. As the tip moves away from the adatom the spin-diode effect vanishes and the currents become polarized for both $V>0$ and $V<0$. We also find an imbalance between the up and down spin populations in the adatom, which can be tuned by the tip position and the bias. Finally, due to the presence of the adsorbate on the surface, we observe spin-resolved Friedel oscillations in the current, which reflects the oscillations in the calculated LDOS of the subsystem surface$+$adatom.
\end{abstract}

\maketitle

\section{Introduction}

The Scanning Tunneling Microscope (STM) has allowed huge advances in condensed matter 
physics. On one hand it serves as a powerful tool to manipulate matter on a single atomic scale;\cite{rb04,pa93} on the other
it is used as a probe to the topology of metallic and semiconductor surfaces.\cite{jcc93} 
An impressive early example of such control is the quantum corral, assembled by moving atom-by-atom on a
metallic surface. \cite{mfc93,hcm00} 

In the fascinating field of spintronics,\cite{dda07} STM  was recently used to manipulate individual
Co atoms adsorbed on a template of Mn.\cite{ds10} It was possible, for instance, to determine the spin direction of the individual Co atoms. STM was also applied to study the interactions between isolated Mn acceptors and the influence of the surface
on the impurity properties in diluted magnetic semiconductors, e.g. Mn-doped GaAs.\cite{cc10,dk06}
More interesting, spin-polarized STM, sensitive to surface magnetization,\cite{ars06} has been used to map 
the morphology and the density of states of single magnetic structures\cite{ho10} and magnetic quantum dots.\cite{nod08}
STM has also been employed in the investigation and identification of promising molecular switches, which could be used in future nanoscale circuits.\cite{mk09,fm10,gs09} In the context of quantum information,\cite{br10} STM was used to measure electron spin relaxation times of individual atoms adsorbed on a surface with nanosecond time resolution. \cite{sl10,aa10}
More recently a new type of Scanning Probe Microscope was demonstrated using ultracold atoms.\cite{mg11} Interestingly, the conventional solid tip is replaced by a gas of ultracold rubidium atoms, which increases the spatial resolution of the microscope. 
All these applications highlight the importance of STM to the development of nano-engineered systems for spintronics and spin-based quantum information processing. 

\begin{figure}[t]
\par
\begin{center}
\epsfig{file=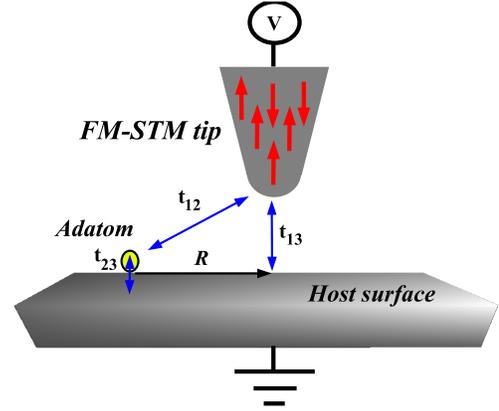, width=0.45\textwidth}
\end{center}
\caption{Ferromagnetic Scanning Tunneling Microscope (FM-STM) tip coupled to a host surface with an adatom. The matrix elements $t_{12}$, $t_{13}$ and $t_{23}$ represent the couplings tip-adatom, tip-surface and adatom-surface, respectively. The tip-adatom lateral distance is denoted by $R$.} \label{fig1}
\end{figure}

As for strong correlated systems, STM has revealed new facets of the Kondo physics, 
such as the Fano-Kondo resonance on the differential conductance when a single magnetic adatom (e.g., Co) is 
placed on a metallic surface.\cite{hcm00,vm98} If the tip is ferromagnetic the Fano-Kondo line shape becomes spin-dependent\cite{krp07} and the setup can be used as a powerful spin filter.\cite{acs09_2}

Here we study spin-dependent transport in a system composed of a ferromagnetic (FM) STM tip coupled
to both an adsorbed atom and a host non-magnetic (NM) surface. This geometry resembles a junction NM-QD-FM, where ``QD'' stands for quantum dot.
In the present system, however, we have one additional ingredient: the tip can move laterally away from the
adatom. It is well known that the NM-QD-FM system gives rise to unpolarized current for direct bias ($V>0$)
and polarized current for reverse bias ($V<0$) when the dot is singly occupied (Coulomb blockade forbids double occupancy).\cite{fms07} 
This rectification of the current polarization is the so called spin-diode effect. In this work
we investigate the interplay between the Coulomb blockade and the lateral displacement of the tip on the spin-diode effect.

Our calculation is based on the Keldysh nonequilibrium technique. By performing a self-consistent calculation we determine the adatom occupation 
and its magnetization as a function of the tip lateral position $R$, Fig. \ref{fig1}. We find that the adatom becomes magnetized when 
the tip approaches it; this magnetization switches sign as the bias is reversed. 
Moreover, we calculate the spin-resolved current in both single and double occupancy regimes of the adatom and find that it can depend strongly on the tip position. In particular, the spin-diode
effect emerges when the tip is closer to the adatom and the charging energy is large enough to
allow for only a single electron in the adatom. As the tip moves away from the adatom the currents become equally polarized for both biases thus resulting in a suppression of the spin-diode behavior. More interesting, we observe spin-resolved Friedel oscillations\cite{pa94,kk01,ia08,acs11,sk10} in the current as the lateral distance tip-adatom $R$ increases. These are due to the presence of the adatom on the surface, and reflect the oscillations in the local density of states (LDOS) of the subsystem surface$+$adatom.

The paper is organized as follows. In Sec. II we present a detailed description of the formulation used 
to compute the spin-resolved currents and the spin populations. We divide this section into three subsections: in A we determine an expression for the spin-resolved currents and the local density of states of the subsystem adatom-surface when the tip is far away from the adatom. In B, the calculation for the current is performed in the presence of the tip, and in C we summarize the numerical technique and the parameters used in the simulations. We present our results and discussions in Sec. III. Section IV summarizes the main ideas of our work.

\section{Formulation}
 
The system we study is composed of a FM-STM tip over an adsorbate on a host surface, Fig. \ref{fig1}. Electrons can tunnel from the tip to the substrate or vice-versa via two possible ways: (i) direct tunneling tip-to-surface or (ii) tunneling via the localized state of the adatom. The system Hamiltonian is
\begin{equation}
 \label{Hamiltonian}
 H=H_1+H_2+H_3+H_{12}+H_{13}+H_{23},
\end{equation}
where $H_i$ corresponds to the tip ($i=1$), the adatom ($i=2$) or the surface ($i=3$), and reads
\begin{equation}\label{Hi}
 H_i = \sum_{\mathbf{k}_i\sigma} \epsilon_{\mathbf{k}_i\sigma} c_{\mathbf{k}_i\sigma}^\dagger c_{\mathbf{k}_i\sigma}+\delta_{i 2}U \hat{n}_\uparrow \hat{n}_\downarrow,
\end{equation} 
where for $i=1(3)$, $\mathbf{k}_i$ is the wave number for electrons in the tip (host) and the label $\s$ stands for the electron spin components $\uparrow$ and $\downarrow$. Here $\epsilon_{\mathbf{k}_i\sigma}$ is the energy of the state $\mathbf{k}_i\sigma$ and $c_{\mathbf{k}_i\sigma}$ ($c^{\dagger}_{\mathbf{k}_i\sigma}$) annihilates (creates) an electron in the quantum state $\mathbf{k}_i\sigma$. We consider a Stoner-like ferromagnetic dispersion $\epsilon_{\mathbf{k}_1\sigma}=\hbar^2k_1^2/2m + \sigma\Delta$ for the tip, with $m$ being the free electron mass and $\Delta$ the usual Stoner parameter,\cite{ec39,fms04} and a free electron dispersion $\epsilon_{\mathbf{k}_3}=\hbar^2k_3^2/2m$ for the surface. For the adatom, $i=2$, we consider only a single spin-degenerate energy level, $\epsilon_{\mathbf{k}_2\sigma}=\epsilon_{\sigma}$. In this case the index $\mathbf{k}_2$ simply denotes the adatom level. The second term in $H_2$ accounts for the Coulomb interaction  $U$ in the adatom. 

The coupling terms in Eq. (\ref{Hamiltonian}) can be written as
\begin{equation}\label{Hij}
 H_{ij}= \sum_{\mathbf{k}_i \mathbf{k}_j \sigma} (t_{ij} c_{\mathbf{k}_i\sigma}^\dagger c_{\mathbf{k}_j\sigma} + t_{ij}^* c_{\mathbf{k}_j\sigma}^\dagger c_{\mathbf{k}_i\sigma}),
\end{equation}
where $t_{ij}$ is the coupling parameter between subsystems $i$ and $j$; $t_{12}$, $t_{13}$ and $t_{23}$ account for the tunnelings tip-adatom, tip-surface and adatom-surface, respectively. When a bias voltage is applied these transfer terms drive the system out of equilibrium.

Next we consider a real space formulation for the spin-resolved current. This is particularly convenient since we are interested in looking at Friedel oscillations on the surface. As we shall see later on, this formulation is equivalent to a formulation in the \textbf{k} space. 

\subsection{Non-resonant transport}
\label{NRT}
For simplicity let us first consider the transport regime in which the direct coupling between
the tip and the adatom is negligible (non-resonant transport), which is valid for large enough tip-adatom lateral distances. 
The Hamiltonian of the system in this case reduces to
\begin{equation}\label{Hfirst}
H=H_1+H_2+H_3+H_{13}+H_{23}.
\end{equation}

The electrical current for spin $\sigma$ between the tip and the surface can be calculated from the definition\cite{hh96}
\begin{equation}\label{Idefinition}
 I_1^\sigma= - e \langle \dot{N}_1^\sigma \rangle=- i e \langle [H,N_1^\sigma] \rangle,
\end{equation}
with $e$ the electron charge ($e>0$) and $N_1^\sigma$ the total number operator given by
\begin{equation}
 N_1^\sigma= \int d\mathbf{r}_1 \Psi^{\sigma^\dagger}_1 (\mathbf{r}_1,t) \Psi^{\sigma}_1 (\mathbf{r}_1,t),
\end{equation}
where $ \Psi^{\sigma}_1 (\mathbf{r}_1,t)$ and $\Psi^{\sigma^\dagger}_1 (\mathbf{r}_1,t)$ are quantum field operators
for the electrons in the tip. In Eq. (\ref{Idefinition}) and throughout the paper we assume $\hbar=1$. 

The quantity $\langle O(t)\rangle$ defines the nonequilibrium average value of a physical observable denoted by the operator $O(t)$, and it is given by\cite{jr86,hh96}
\begin{equation}
\langle O(t) \rangle=\mbox{Tr}[\r O(t)],
\end{equation}
where $\r$ is the thermal equilibrium density matrix, $\r=(\mbox{Tr}e^{-\beta H_0})^{-1}e^{-\beta H_0}$, with $H_0$ being the Hamiltonian containing only the $H_i$ terms in Eq. (\ref{Hfirst}), and $O(t)$ is in the Heisenberg picture, i.e., its time-dependence is governed by the full Hamiltonian of Eq. (\ref{Hfirst}).

The only non-vanishing term in the commutator of Eq. (\ref{Idefinition}) is $[H_{13},N_1^\sigma]$. For electrons with spin $\sigma$, the tip-surface coupling can be written as\cite{bf}
\begin{equation}\label{H13real}
 H_{13}=\sum_{\sigma}\int \int d\mathbf{r}_1 d\mathbf{r}_3 [T(\mathbf{r}_1,\mathbf{r}_3) \Psi^{\sigma^\dagger}_1 (\mathbf{r}_1,t) \Psi^{\sigma}_3 (\mathbf{r}_3,t)+h.c.],
\end{equation}
where $T(\mathbf{r}_1,\mathbf{r}_3)$ is a matrix element that accounts for the coupling between the tip and the surface, and $\Psi^{\sigma}_3 (\mathbf{r}_3,t)$ is the quantum field
operator for electrons in the surface. Calculating $[H_{13},N_1^\s]$ and using the result in Eq. (\ref{Idefinition}) we find for the spin-resolved current
\begin{eqnarray}
 I_1^{\sigma} &=& i e \int \int d\mathbf{r}_1 d\mathbf{r}_3 [T(\mathbf{r}_1,\mathbf{r}_3)  \langle \Psi^{\sigma^\dagger}_1 (\mathbf{r}_1,t) \Psi^{\sigma}_3 (\mathbf{r}_3,t)\rangle  - \nonumber \\ && 
\phantom{xxxxxxxxxxx} T^*(\mathbf{r}_1,\mathbf{r}_3)  \langle \Psi^{\sigma^\dagger}_3 (\mathbf{r}_3,t) \Psi^{\sigma}_1 (\mathbf{r}_1,t)\rangle].
\end{eqnarray}
Defining the lesser Green function
\begin{eqnarray}
 G_{\sigma}^<(\mathbf{r}_3,t_3;\mathbf{r}_1,t_1)&=&i \langle \Psi^{\sigma^\dagger}_1 (\mathbf{r}_1,t_1) \Psi^{\sigma}_3 (\mathbf{r}_3,t_3)\rangle,
\end{eqnarray}
we can rewrite the current as
\begin{equation}\label{I1meio}
  I_1^{\sigma} = 2 e \mathrm{Re} \left\{ \int \int d\mathbf{r}_1 d\mathbf{r}_3 T(\mathbf{r}_1,\mathbf{r}_3) G_{\sigma}^<(\mathbf{r}_3,t;\mathbf{r}_1,t) \right\}.
\end{equation}

We now aim at determining $G_{\sigma}^<(\textbf{r}_3,t;\textbf{r}_1,t)$ in Eq. ({\ref{I1meio}}). To this end, we use the nonequilibrium Keldysh formalism. Similarly to the equilibrium case, here we introduce an ordered Green function 
\begin{equation}
\label{cgf}
G_{\sigma}(\mathbf{r}_3,\tau_3;\mathbf{r}_1,\tau_1)=-i\langle T_C \Psi^{\sigma}_3 (\mathbf{r}_3,\tau_3) \Psi^{\sigma^\dagger}_1 (\mathbf{r}_1,\tau_1)\rangle,
\end{equation}
with the $\tau$'s defined, however, on a contour $C$ in the complex plane. The operator $T_C$, called contour-ordering operator, orders the operators according to the position of their time arguments on the contour. From the contour-ordered Green function we can obtain the lesser $G_{\sigma}^<$, greater $G_{\sigma}^>$, retarded $G_{\sigma}^r$ and advanced $G_{\sigma}^a$ Green functions, which are directly linked to the observables.

To obtain $G_{\sigma}^<$, $G_{\sigma}^>$, $G_{\sigma}^r$ and $G_{\sigma}^a$, we first determine the equation of motion for the Green function in Eq. (\ref{cgf}), 
\begin{eqnarray}
\label{em1}
\left(i\frac{\partial}{\partial \tau_1}-\frac{\nabla^2}{2m}\right)G_{\sigma}(\mathbf{r}_3,\tau_3;\mathbf{r}_1,\tau_1)&=&-\int d\mathbf{r}'_3 T(\mathbf{r}'_3,\mathbf{r}_1) \times \nonumber \\ 
& &  G_{\sigma}(\mathbf{r}_3,\tau_3;\mathbf{r}'_3,\tau'_3), \nonumber \\
\end{eqnarray}
or in the integral form
\begin{eqnarray}
\label{em2}
G_{\sigma}(\mathbf{r}_3,\tau_3;\mathbf{r}_1,\tau_1)&=&\int \int d\mathbf{r}'_1 d\mathbf{r}'_3 \int_C d\tilde{\tau}G_{\sigma}(\mathbf{r}_3,\tau_3;\mathbf{r}'_3,\tilde{\tau})\times \nonumber \\
 & & \phantom{xxxxxx}T(\mathbf{r}'_3,\mathbf{r}'_1)g_{\sigma}(\mathbf{r}'_1,\tilde{\tau};\mathbf{r}_1,\tau_1),
\end{eqnarray}
where $g_{\sigma}$ is the free-electron Green function of the tip and the time integral is over the contour $C$; then we perform an appropriate analytical continuation. This procedure consists essentially in replacing the contour integral over $\tau$ in Eq. (\ref{em1}) by a real time  integral over $t$. Here we follow the Langreth procedure. \cite{dl76}
For the lesser Green function $G_{\sigma}^<$ we have  
\begin{eqnarray}\label{Glesser1}
 && G_{\sigma}^<(\mathbf{r}_3,t_3;\mathbf{r}_1,t_1)=\int \int d\mathbf{r}'_1 d\mathbf{r}'_3 \int d\tilde{t} \times \nonumber \\ && 
\left[G_{\sigma}^r(\mathbf{r}_3,t_3;\mathbf{r}'_3,\tilde{t})  T(\mathbf{r}'_3,\mathbf{r}'_1) g_{\sigma}^{<}(\mathbf{r}'_1,\tilde{t};\mathbf{r}_1,t_1)+ \right.\nonumber \\ &&
\left. G_{\sigma}^<(\mathbf{r}_3,t_3;\mathbf{r}'_3,\tilde{t})  T(\mathbf{r}'_3,\mathbf{r}'_1) g_{\sigma}^{a}(\mathbf{r}'_1,\tilde{t};\mathbf{r}_1,t_1)\right].
\end{eqnarray}
In the above equation $g_{\sigma}^a$ and $g_{\sigma}^<$ correspond to the analytically continued free-electron advanced and lesser Green functions of the tip, respectively. Throughout the paper we use lower case to denote the free-electron Green functions of the tip, the adatom and the surface. We note that $G_{\sigma}^<(\mathbf{r}_3,t_3;\mathbf{r}_1,t_1)$ is coupled to $G_{\sigma}^r(\mathbf{r}_3,t_3;\mathbf{r}'_3,\tilde{t})$ and also to $G_{\sigma}^<(\mathbf{r}_3,t_3;\mathbf{r}'_3,\tilde{t})$. To completely determine $G_{\sigma}^<(\mathbf{r}_3,t_3;\mathbf{r}_1,t_1)$ we then need to perform an iterative process and obtain a system of equations for the Green functions $G_{\sigma}^r$ and $G_{\sigma}^<$.

Substituting Eq. (\ref{Glesser1}) into Eq. (\ref{I1meio}) we obtain
\begin{eqnarray}
  I_1^{\sigma} &=& 2 e \mathrm{Re} \Biggl\{ \int \int \int \int \int d\mathbf{r}_1 d\mathbf{r}_3 d\mathbf{r}'_1 d\mathbf{r}'_3 d\tilde{t} T(\mathbf{r}_1,\mathbf{r}_3) T(\mathbf{r}'_3,\mathbf{r}'_1) \times \nonumber \\ && 
 \phantom{xxxxxx}\left[G_{\sigma}^r(\mathbf{r}_3,t;\mathbf{r}'_3,\tilde{t}) g_{\sigma}^{<}(\mathbf{r}'_1,\tilde{t};\mathbf{r}_1,t) \right. +\nonumber \\ &&
 \phantom{xxxxxxxxx} \left. G_{\sigma}^<(\mathbf{r}_3,t;\mathbf{r}'_3,\tilde{t}) g_{\sigma}^{a}(\mathbf{r}'_1,\tilde{t};\mathbf{r}_1,t)\right] \Biggr\}.
\end{eqnarray}
Performing a Fourier transform in the time coordinate we find
\begin{eqnarray}
  I_1^{\s} &=& 2 e \int \frac{d\omega}{2\pi}\mathrm{Re} \Biggl\{ \int \int  \int \int d\mathbf{r}_1 d\mathbf{r}_3  d\mathbf{r}'_1 d\mathbf{r}'_3  \times \\&& \phantom{xx}T(\mathbf{r}_1,\mathbf{r}_3) T(\mathbf{r}'_3,\mathbf{r}'_1)[G_{\s}^r(\mathbf{r}_3,\mathbf{r}'_3,\omega) g_{\s}^{<}(\mathbf{r}'_1,\mathbf{r}_1,\omega)+  \nonumber \\ && 
\phantom{xxxxxxxxxxxxxxxx}  G_{\s}^<(\mathbf{r}_3,\mathbf{r}'_3,\omega) g_{\sigma}^{a}(\mathbf{r}'_1,\mathbf{r}_1,\omega)] \Biggr\}.\nonumber
\end{eqnarray}
Additionally, Fourier transforming $g_{\s}^{<}$ and $g_{\s}^{a}$ results in
\begin{eqnarray}
  && I_1^{\s} = 2 e \int \frac{d\omega}{2\pi} \mathrm{Re} \Biggl\{ \sum_{\mathbf{k}_1} \int \int  \int \int d\mathbf{r}_1 d\mathbf{r}_3  d\mathbf{r}'_1 d\mathbf{r}'_3 \times \nonumber  \\ 
&& e^{-i \mathbf{k}_1 \cdot \mathbf{r}_1} T(\mathbf{r}_1,\mathbf{r}_3)  e^{i \mathbf{k}_1 \cdot \mathbf{r}'_1} T(\mathbf{r}'_3,\mathbf{r}'_1)   \\ && 
  [G_{\s}^r(\mathbf{r}_3,\mathbf{r}'_3,\omega) g_{\mathbf{k_1}\s}^{<}(\omega)+
G_{\s}^<(\mathbf{r}_3,\mathbf{r}'_3,\omega) g_{\mathbf{k}_1\s}^{a}(\omega)] \Biggr\}.\nonumber
\end{eqnarray}
Now we assume that the coupling between the tip and the surface is local \cite{bf,gf03} (point source), i.e., 
$T(\mathbf{r}_1,\mathbf{r}_3) = T^0(\mathbf{r}_1) \delta(\mathbf{r}_3-\mathbf{R})$, $T^0(\mathbf{r}_1)$ being a proportionality factor.
Integrating over the surface coordinates we find
\begin{eqnarray}
  && I_1^{\s} = 2 e \int \frac{d\omega}{2\pi} \mathrm{Re} \Biggl\{ \sum_{\mathbf{k}_1}  \int \int d\mathbf{r}_1  d\mathbf{r}'_1 e^{-i \mathbf{k}_1 \cdot \mathbf{r}_1}  T^0(\mathbf{r}_1)\times   \nonumber  \\ && 
 \phantom{xxxx}e^{i \mathbf{k}_1 \cdot \mathbf{r}'_1} T^{0^*}(\mathbf{r}'_1)[G_{\s}^r(\mathbf{R},\mathbf{R},\omega) g_{\mathbf{k}_1\s}^{<}(\omega) + \nonumber \\ &&
\phantom{xxxxxxxxxxxxxxx}  G_{\s}^<(\mathbf{R},\mathbf{R},\omega) g_{\mathbf{k}_1\s}^{a}(\omega)] \Biggr\},
\end{eqnarray}
and then over $\mathbf{r}_1$ and $\mathbf{r}'_1$ we arrive at
\begin{eqnarray}\label{I1final1}
  && I_1^{\s} = 2 e \int \frac{d\omega}{2\pi} \mathrm{Re} \Biggl\{ \sum_{\mathbf{k}_1} |t^0_{13\mathbf{k_1}}|^2 \times  \\ && 
 [G_{\s}^r(\mathbf{R},\mathbf{R},\omega) g_{\mathbf{k}_1\s}^{<}(\omega)+
G_{\s}^<(\mathbf{R},\mathbf{R},\omega) g_{\mathbf{k}_1\s}^{a}(\omega)] \Biggr\}, \nonumber
\end{eqnarray}
where $t^0_{13\mathbf{k_1}}$ is the Fourier transform of $T^0(\mathbf{r}_1)$. It is convenient to perform a Fourier transform on $G_{\s}^r(\mathbf{R},\mathbf{R},\w)$ and $G_{\s}^<(\mathbf{R},\mathbf{R},\w)$. This results in 
\begin{eqnarray}\label{curr1}
  && I_1^{\s} = 2 e \int \frac{d\omega}{2\pi} \mathrm{Re} \Biggl\{ \sum_{\mathbf{k}_3,\mathbf{k}'_3} \sum_{\mathbf{k}_1}  
 t_{13\mathbf{k}_1\mathbf{k}_3} t_{13\mathbf{k}_1\mathbf{k}'_3}^*  \times\\ && 
 [G_{\mathbf{k}_3 \mathbf{k}'_3,\s}^r(\omega) g_{\mathbf{k}_1\s}^{<}(\omega)+
G_{\mathbf{k}_3 \mathbf{k}'_3,\s}^<(\omega) g_{\mathbf{k}_1\s}^{a}(\omega)] \Biggr\}, \nonumber
\end{eqnarray}
where $t_{13\mathbf{k}_1\mathbf{k}_3}=t^0_{13\mathbf{k}_1} e^{i\mathbf{k}_3 \cdot \mathbf{R}}$.\cite{comment1}
In what follows we assume $t^0_{13\mathbf{k}_1}$ to be a constant, i.e., independent of $\mathbf{k_1}$. As we have mentioned before, Eq. (\ref{curr1}) could have been directly derived from the $\mathbf{k}$ space Hamiltonian 
$H_{13}^\s=\sum_{\mathbf{k}_1,\mathbf{k}_3}  t_{13} c_{\mathbf{k}_1\s}^\dagger c_{\mathbf{k}_3\s} + 
t_{13}^* c_{\mathbf{k}_3\s}^\dagger c_{\mathbf{k}_1\s}$, with $t_{13}=t_{13}^0 e^{i\mathbf{k}_3 \cdot \mathbf{R}}$. Observe that this 
is completely equivalent to the real space Hamiltonian Eq. (\ref{H13real}), with a point source,
$T(\mathbf{r}_1,\mathbf{r}_3) = T^0(\mathbf{r}_1) \delta(\mathbf{r}_3-\mathbf{R})$. 

\subsubsection{Spatially resolved transmission coefficient} 
\label{NRT1}
Equation (\ref{I1final1}) can alternatively be written as
\begin{eqnarray}
  && I_1^{\s} = e \int \frac{d\w}{2\pi} \left[2 \pi |t^0_{13}|^2 \rho_{1\sigma}(\w)\right]i \left\{G_{\s}^<(\mathbf{R},\mathbf{R},\w) \right. + \\ && 
  \phantom{xxxxxxxx} \left. f_1(\w) [G_{\s}^r(\mathbf{R},\mathbf{R},\w)-G_{\s}^a(\mathbf{R},\mathbf{R},\w)] \right\}, \nonumber
\end{eqnarray}
where $f_1(\w)$ and $\rho_{1\s}(\w)=\sum_{\mathbf{k}_1}  \delta(\w-\epsilon_{\mathbf{k}_1\s})$ are, respectively, the Fermi function and the density of states of the tip.
If the tip is weakly coupled to the surface, we can calculate the current in the lowest order
of $|t^0_{13}|^2$. This means that the Green functions $G_{\s}^r(\mathbf{R},\mathbf{R},\w)$ and $G_{\s}^<(\mathbf{R},\mathbf{R},\w)$
are assumed tip-decoupled, i.e., here they are local Green functions for the subsystem adatom-surface only.
In other words, the tip behaves as a probe, not affecting the local density of states.
This allows us to write the current as
\begin{equation}\label{I1T}
 I_1^{\s}=e \int \frac{d\w}{2\pi} \left[f_1(\w)-f_3(\w)\right] T_{\s}(\mathbf{R},\w),
\end{equation}
where $f_3(\w)$ is the Fermi function of the surface and the transmission coefficient is defined by
\begin{eqnarray}\label{transmission}
  T_{\s}(\mathbf{R},\w) &=& \left[2 \pi |t^0_{13}|^2 \rho_{1\sigma}(\w)\right]\left[-2\mathrm{Im}\{G_{\s}^r(\mathbf{R},\mathbf{R},\w)\}\right] \nonumber \\
                   &=& \left[2 \pi |t^0_{13}|^2 \rho_{1\sigma}(\w)\right]\left[2\pi \mathrm{\rho}_{\s}(\mathbf{R},\w)\right],
\end{eqnarray}
with $\rho_{\s}\left(\mathbf{R},\w\right)=-\frac{1}{\pi} \mathrm{Im}\{G_{\s}^r(\mathbf{R},\mathbf{R},\w)\}$ -- the local density of states (LDOS).
In particular, in the absence of the adatom, the LDOS becomes the unperturbed surface density of states per spin, $\rho_{3}(\w)=\sum_{\mathbf{k}_3}  \delta(\w-\epsilon_{\mathbf{k}_3})=m/n\pi \hbar^2=1/2D=\rho_3^0$, with $n$ the 2D electronic density and $D$ the band half width. 

In this case and in the zero temperature limit, the current reads
\begin{equation}\label{Ianalytic}
 I_1^{\s}=e 2 \pi |t^0_{13}|^2 \rho_{1\sigma} \rho_3^0 eV,
\end{equation}
where $V$ is the bias voltage. 

As we shall see in the next section, in the presence of the adatom, similarly to the case of an impurity on the surface of a metal, the LDOS shows Friedel oscillations; these affect the current that oscillates around the value given by Eq. (\ref{Ianalytic}).\cite{comment2}

\subsubsection{Calculating $\rho_{\sigma}(\mathbf{R},\w)$ for the subsystem adatom+surface}

To determine the LDOS in the presence of the adatom, we calculate $G_{\s}^r(\mathbf{r}_3,\mathbf{r}'_3,\w)$ by Fourier transforming $G_{\mathbf{k}_3 \mathbf{k}'_3,\s}^r(\w)$, 
\begin{equation}\label{GrFT}
 G_{\s}^r(\mathbf{r}_3,\mathbf{r}'_3,\w)=\sum_{\mathbf{k}_3 \mathbf{k}'_3} e^{i\mathbf{k}_3 \cdot \mathbf{r}_3} e^{-i\mathbf{k}'_3\cdot \mathbf{r}'_3} G_{\mathbf{k}_3 \mathbf{k}'_3,\s}^r(\w),
\end{equation}
assuming there is no tip-to-surface coupling, i.e., considering $H=H_2+H_3+H_{23}$. Following the procedure we described in the previous section -- equation of motion + analytical continuation -- we obtain\cite{comment3}
\begin{eqnarray}\label{GrFT2}
  G_{\mathbf{k}_3 \mathbf{k}'_3,\s}^r(\w)&=&\delta_{\mathbf{k}_3,\mathbf{k}'_3}g_{\mathbf{k}_3\s}^r(\w)+ \sum_{\mathbf{k}_2 \mathbf{k}'_2}|t_{23}|^2 \times \nonumber \\ & & g_{\mathbf{k}_3\s}^r(\w)g_{\mathbf{k}'_3\s}^r(\w) G_{\mathbf{k}_2 \mathbf{k}'_2,\s}^r(\w),
\end{eqnarray}
where $g_{\mathbf{k}_3\s}^r(\w)=(\w-\epsilon_{k_3}+i \delta)^{-1}$ and $\sum_{\mathbf{k}_2 \mathbf{k}'_2} G_{\mathbf{k}_2 \mathbf{k}'_2,\s}^r(\w)\equiv G_{22\s}^r(\w)$ is the adatom retarded Green function. 
Using Eq. (\ref{GrFT2}) in Eq. (\ref{GrFT}) we find
\begin{eqnarray}
 && G_{\s}^r(\mathbf{r}_3,\mathbf{r}'_3,\w) = \sum_{\mathbf{k}_3} \frac{e^{i\mathbf{k_3}\cdot(\mathbf{r_3}-\mathbf{r'_3})}}{\w-\epsilon_{k_3}+i \delta} + |t_{23}|^2 \times \nonumber \\ && \phantom{xxxxx}
 \sum_{\mathbf{k}_3} \frac{e^{i\mathbf{k}_3 \cdot \mathbf{r}_3}}{\w-\epsilon_{k_3}+i \delta}
 \sum_{\mathbf{k}'_3} \frac{e^{-i\mathbf{k}'_3 \cdot \mathbf{r}'_3}}{\w-\epsilon_{k'_3}+i \delta} G_{22\s}^r(\w).
\end{eqnarray}

Let
\begin{equation}
 \sum_{\mathbf{k}_3} \frac{e^{i\mathbf{k}_3 \cdot \mathbf{r}_3}}{\w-\epsilon_{k_3}+i \delta}=R(r_3,\w)+i I(r_3,\w),
\end{equation}
where $R(r_3,\w)$ and $I(r_3,\w)$ denote the corresponding real and imaginary parts given by
\begin{equation}
 R(r_3,\w)=\rho_{3}^0 \int_{-1}^{1} dx \frac{\frac{\w}{D}-x}{\left(\frac{\w}{D}-x\right)^2+\delta^2} J_0\left(k_F r_3 \sqrt{1+x}\right),
\end{equation}
and 
\begin{equation}
 I(r_3,\w)=-\pi \rho_3^0 J_0\left(k_F r_3 \sqrt{1+\frac{\w}{D}}\right), 
\end{equation}
with $k_F$ being the Fermi wave number.
We can then write the LDOS as
\begin{eqnarray}
 &&\rho_{\s}(R,\w)=\rho_3^0 \Biggl\{ 1 + \frac{\Gamma_{3}}{2} J_0^2\left(k_F R \sqrt{1+\frac{\w}{D}}\right) \times \nonumber \\ && \phantom{xxxxxxxxxxx} 
 \left[(1-q^2)\mathrm{Im}G_{22\s}^r-2q\mathrm{Re}G_{22\s}^r(\w)\right]\Biggr\}, \nonumber \\
\end{eqnarray}
where $\Gamma_3=2\pi|t_{23}|^2\rho_3^0$ and $q=R(R,\w)/I(R,\w)$. 
In order to determine the transmission coefficient, we only have to calculate the adatom retarded Green function $G_{22\s}^r(\w)$, obtained here using the Hubbard I approximation.\cite{hh96} 
This approximation accounts for the Coulomb interaction and consists in factorizing the higher-order correlation functions appearing in the resulting equation of motion for $G_{22\s}^r(\w)$. As a result we have\cite{hh96}
\begin{equation}
  G_{22\s}^r(\w)=\frac{1} {{g}_{2\s}^{r^{-1}}(\w)-\S^r(\w)},
 \end{equation}
with
\begin{eqnarray}\label{gradatom}
 g_{2\s}^r(\w) = \frac{\w-\epsilon_\s-U(1-n_{\bar{\s}})}{(\w-\epsilon_\s)(\w-\epsilon_\s-U)},
\end{eqnarray}
where $\bar{\sigma}=-\s$, $n_{\bar{\s}}$ is the average occupation and $\S^r$ is the self energy related to the coupling between the adatom and the host surface, $\S^r=-\frac{i}{2}\Gamma_3$.

\subsection{Resonant + non-resonant transport}

In this section we determine the \textit{total} current -- tip-to-adatom-to-host + tip-to-host -- flowing in the system. In addition to the non-resonant (tip-to-host) current, calculated in Section A, here we consider the contribution from the direct tunneling of electrons between the tip and the adatom (resonant current). The corresponding additional term $H_{12}$ is now taken into account and the Hamiltonian describing 
the system is given by Eq. (\ref{Hamiltonian}). Note that for large enough $R$ distances ($R \rightarrow \infty$) we regain the non-resonant case. Here for convenience we perform the calculation in {\bf{k}} space. 

We model the tip-surface, tip-adatom 
and adatom-surface couplings, respectively, by considering 
\begin{eqnarray}
 t_{12}&=&t_{12}^0e^{-R/R_0},\label{t12}\\
 t_{13}&=&t_{13}^0 e^{i\mathbf{k}_3 \cdot \mathbf{R}},\label{t13}\\ 
 t_{23}&=&t_{23}^0,\label{t23}
\end{eqnarray}
where $t_{12}^0$, $t_{13}^0$ and $t_{23}^0$ are (constant) phenomenological parameters
and $R_0$ gives the exponential spatial decay for the coupling between the tip and the adatom as the
tip moves away from it.

The current flowing into the tip or leaving it can be determined from Eq. (\ref{Idefinition}). Calculating the commutator in this equation via Eqs. (\ref{Hamiltonian})-(\ref{Hij}) we find
\begin{eqnarray}
 [H,N_1^{\s}]&=&\sum_{j=2}^3 [H_{1j},N_1^{\s}] \nonumber \\ &=& \sum_{j=2}^3 \sum_{\mathbf{k}_1,\mathbf{k}_j} \left(-t_{1 j} c_{\mathbf{k}_1\s}^\dagger c_{\mathbf{k}_j\s} +  t_{1 j}^* c_{\mathbf{k}_j\s}^\dagger c_{\mathbf{k}_1\s} \right).
\end{eqnarray}
Substituting this result into Eq. (\ref{Idefinition}) we obtain
\begin{eqnarray}
 I_1^{\s}=-i e \sum_{j=2}^3 \sum_{\mathbf{k}_1, \mathbf{k}_j} \left\{-t_{1j} \langle c_{\mathbf{k}_1\s}^\dagger c_{\mathbf{k}_j\s} \rangle + t_{1j}^* \langle c_{\mathbf{k}_j\s}^\dagger c_{\mathbf{k}_1\s} \rangle \right\},\nonumber \\
\end{eqnarray}
or in terms of the lesser Green function $G_{\mathbf{k}_j \mathbf{k}_1, \s}^<(t,t)$,
\begin{equation}\label{I1Glesser}
 I_1^{\s} = 2 e \mathrm{Re} \left\{ \sum_{j=2}^3 \sum_{\mathbf{k}_1, \mathbf{k}_j} t_{1j} G_{\mathbf{k}_j \mathbf{k}_1, \s}^<(t,t) \right\},
\end{equation}
where $G_{\mathbf{k}_j \mathbf{k}_1, \s}^<(t,t)=i\langle c_{\mathbf{k}_1\s}^\dagger(t) c_{\mathbf{k}_j\s}(t) \rangle$. Equation (\ref{I1Glesser}) is equivalent to Eq. (\ref{I1meio}) when $t_{12}=0$. Now we must find an expression for the lesser Green function. Following the procedure described in section A, below we write down the equation of motion for the contour-ordered Green function $G_{\mathbf{k}_j \mathbf{k}_1,\s}(\tau,\tau')=-i\langle T_C c_{\mathbf{k}_j\s}(\tau) c_{\mathbf{k}_1\s}^\dagger(\tau') \rangle$ 
\begin{equation}
\left(i\frac{\partial}{\partial \tau'}+\epsilon_{\mathbf{k}_1}\right)G_{\mathbf{k}_j \mathbf{k}_1,\s}(\tau,\tau')=-\sum_{l=2}^3 \sum_{\mathbf{k}_l} t_{1 l}^* G_{\mathbf{k}_j \mathbf{k}_l,\s}(\tau,\tau'),
\end{equation}
or in the integral form
\begin{equation}\label{GF1}
 G_{\mathbf{k}_j \mathbf{k}_1,\s}(\tau,\tau')=\sum_{l=2}^3 \sum_{\mathbf{k}_l} t_{1 l}^* \int_C d\tau_1 G_{\mathbf{k}_j \mathbf{k}_l,\s}(\tau,\tau_1) g_{\mathbf{k}_1\s}(\tau_1,\tau'),
\end{equation}
where $g_{\mathbf{k}_1\s}(\tau_1,\tau')$ is the tip free-electron Green function, and then analytically continue Eq. (\ref{GF1}) to find $G_{\mathbf{k}_j \mathbf{k}_1,\s}^<(t,t')$. Using this result in Eq. (\ref{I1Glesser}) we have
\begin{eqnarray}
 I_1^{\s} &=& 2 e \mathrm{Re} \Biggl\{ \sum_{j,l=2}^3 \sum_{\mathbf{k}_1, \mathbf{k}_j, \mathbf{k}_l} t_{1j} t_{1 l}^* \int dt_1 \times  \\
     && \left[G_{\mathbf{k}_j \mathbf{k}_l,\s}^r(t,t_1) g_{\mathbf{k}_1\s}^<(t_1,t) + G_{\mathbf{k}_j \mathbf{k}_l,\s}^<(t,t_1) g_{\mathbf{k}_1\s}^a(t_1,t) \right]\Biggr\}. \nonumber
\end{eqnarray}
Taking the Fourier transform of the above expression we find
\begin{eqnarray}\label{I1first}
  &&I_1^{\s} = 2 e \mathrm{Re} \Biggl\{\int \frac{d\w}{2 \pi} \times   \nonumber \\  &&
   \sum_{\mathbf{k}_1 \mathbf{k}_2 \mathbf{k}'_2} t_{12} t^*_{12}  [G^r_{\mathbf{k}_2 \mathbf{k}'_2,\s}(\w) g^<_{\mathbf{k}_1\s}(\w) +  G_{\mathbf{k}_2 \mathbf{k}'_2,\s}^<(\w) g^a_{\mathbf{k}_1\s}(\w) ] + \nonumber \\ &&
   \sum_{\mathbf{k}_1 \mathbf{k}_2 \mathbf{k}_3} t_{12} t_{13\mathbf{k}_3}^* [G^r_{\mathbf{k}_2 \mathbf{k}_3,\s}(\w) g^<_{\mathbf{k}_1\s}(\w)  + G_{\mathbf{k}_2 \mathbf{k}_3,\s}(\w)^< g^a_{\mathbf{k}_1\s}(\w)] + \nonumber \\ &&
   \sum_{\mathbf{k}_1 \mathbf{k}_3 \mathbf{k}_2} t_{13\mathbf{k}_3} t_{12}^* [G^r_{\mathbf{k}_2 \mathbf{k}_3,\s}(\w) g^<_{\mathbf{k}_1\s}(\w)  +  G_{\mathbf{k}_3 \mathbf{k}_2,\s}^<(\w) g^a_{\mathbf{k}_1\s}(\w)] + \nonumber \\ &&
   \sum_{\mathbf{k}_1 \mathbf{k}_3 \mathbf{k}'_3} t_{13\mathbf{k}_3} t^*_{13\mathbf{k}'_3} [G^r_{\mathbf{k}_3 \mathbf{k}'_3,\s}(\w) g^<_{\mathbf{k}_1\s}(\w) + G_{\mathbf{k}_3 \mathbf{k}'_3,\s}^<(\w) g^a_{\mathbf{k}_1\s}(\w)  ]\Biggr\}. \nonumber \\ &&
\end{eqnarray}
Using Eqs. (\ref{t12})-(\ref{t13}) we can rewrite Eq. (\ref{I1first}) as
\begin{eqnarray}\label{I1second}
 I_{1}^\s &=& 2e \mathrm{Re} \Biggl\{ \int \frac{d\w}{2 \pi} |t_{12}^0|^2 e^{-2 (R/R_0)} [G_{22\s}^r g_{1\s}^< + G_{22\s}^< g_{1\s}^a] + \nonumber \\ &&
                                      \phantom{xxxxxxxxxx} t_{12}^{0} t_{13}^{0^*} e^{-R/R_0} [G_{\underline{3}2\s}^r g_{1\s}^<  + G_{\underline{3}2\s}^< g_{1\s}^a ] +\nonumber \\ &&
                                      \phantom{xxxxxxxxxx} t_{13}^{0} t_{12}^{0^*} e^{-R/R_0} [G_{2\underline{3}\s}^r g_{1\s}^<  +        G_{2\underline{3}\s}^< g_{1\s}^a ] +\nonumber \\ &&
                                      \phantom{xxxxxxxxxx} |t_{13}^0|^2 [G_{\underline{3}\underline{3}\s}^r g_{1\s}^<  + G_{\underline{3} \underline{3}\s}^< g_{1\s}^a]\Biggr\},
\end{eqnarray}
where we have introduced the definitions
\begin{eqnarray}
 g^{<,a}_{1\s}(\w)&=&\sum_{\mathbf{k}_1}g_{\mathbf{k}_1\s}^{<,a}(\w),\nonumber \\
 G_{\underline{3}2\s}^{<,r} &=& \sum_{\mathbf{k}_3 \mathbf{k}_2} e^{i\mathbf{k}_3 \cdot \mathbf{R}} G_{\mathbf{k}_3 \mathbf{k}_2, \s}^{<,r}, \nonumber \\ 
 G_{2 \underline{3}\s}^{<,r} &=& \sum_{\mathbf{k}_2 \mathbf{k}_3} e^{-i\mathbf{k}_3 \cdot \mathbf{R}} G_{\mathbf{k}_2 \mathbf{k}_3, \s}^{<,r}, \nonumber \\
 G_{\underline{3} \underline{3}\s}^{<,r} &=& \sum_{\mathbf{k}_3 \mathbf{k}'_3} e^{i(\mathbf{k}_3-\mathbf{k}'_3)\cdot \mathbf{R}} G_{\mathbf{k}_3 \mathbf{k}'_3, \s}^{<,r}. \nonumber \\
\end{eqnarray}
Note that from Eq. (\ref{I1first}) we regain Eq. (\ref{curr1}) in the limit $t_{12}^0=0$, i.e., when the tip is far away from the adatom.

\subsubsection{Matrix Green function formulation}

We can see from Eq. (\ref{I1first}) that $G_{\mathbf{k}_j \mathbf{k}_1,\s}^<(t,t')$ is coupled to other Green functions. In order to find these Green functions, we have to apply the equation of motion
technique to the corresponding contour-ordered Green function for each one of them and then perform an analytical continuation to obtain the respective $G^{<,r}$. After a straightforward calculation we find
\begin{eqnarray}\label{G1j}
 && G_{\mathbf{k}_i \mathbf{k}_j,\s}(\tau,\tau') = \delta_{\mathbf{k}_i \mathbf{k}_j} g_{\mathbf{k}_j\s}(\tau,\tau') + \nonumber \\ && \sum_{l (l \neq j)} \sum_{\mathbf{k}_l} \int d\widetilde{\tau} G_{\mathbf{k}_i,\mathbf{k}_l, \s}(\tau,\widetilde{\tau}) t_{j l}^* g_{\mathbf{k}_j\s}(\widetilde{\tau},\tau'),
\end{eqnarray}
i.e., a system of coupled equations for the Green functions. Here $g_{\mathbf{k}_j\s}(\tau,\tau')$ is the free-electron Green function of the tip ($j=1$), the adatom ($j=2$) or the host surface ($j=3$). 
These three Green functions $g_{\mathbf{k}_j\s}(\tau,\tau')$ ($j=1,2,3$) can be easily evaluated. Interestingly, by defining 
\begin{eqnarray}
 G_{i j \s}&=&{\sum_{\mathbf{k}_i \mathbf{k}_j}} G_{\mathbf{k}_i \mathbf{k}_j,\s},\label{Gij} \\
 G_{\underline{3} j \s} &=& {\sum_{\mathbf{k}_3 \mathbf{k}_j}} e^{i\mathbf{k}_3 \cdot \mathbf{R}} G_{\mathbf{k}_3 \mathbf{k}_j,\s},\label{Gul3j} \\
 G_{j \underline{3} \s} &=& {\sum_{\mathbf{k}_j \mathbf{k}_3}} e^{-i\mathbf{k}_3 \cdot \mathbf{R}} G_{\mathbf{k}_j \mathbf{k}_3, \s}, \label{Gjul3} \\
 G_{\underline{3} \underline{3} \s} &=& {\sum_{\mathbf{k}_3 \mathbf{k}'_3}}  e^{i(\mathbf{k}_3-\mathbf{k}'_3)\cdot \mathbf{R}} G_{\mathbf{k}_3 \mathbf{k}'_3, \s},\label{Gul3ul3}
\end{eqnarray}
where the sum \textit{is not} taken over the spin indices, we can write down a Dyson equation of the form
\begin{equation}\label{dysonmatrix}
 \mathbf{G}_\s(\tau,\tau')=\mathbf{g}_\s(\tau,\tau')+\int  d\tau_1  \mathbf{G}_\s(\tau,\tau_1) \mathbf{\Sigma} \mathbf{g}_\s(\tau_1,\tau'),
\end{equation}
with $\mathbf{G_\s}(\tau,\tau')$ being a matrix Green function whose elements are defined following Eqs. (\ref{Gij})-(\ref{Gul3ul3}), i.e.,
\begin{equation}
  \mathbf{G}_\s=\left(%
\begin{array}{cccc}
 G_{11\s} & G_{12\s} & G_{13\s} & G_{1\underline{3}\s} \\
 G_{21\s} & G_{22\s} & G_{23\s} & G_{2\underline{3}\s} \\
 G_{31\s} & G_{32\s} & G_{33\s} & G_{3\underline{3}\s} \\
 G_{\underline{3}1\s} & G_{\underline{3}2\s} & G_{\underline{3}3\s} & G_{\underline{3}\underline{3}\s}
\end{array} \right).
\end{equation}
Additionally the self-energy is given by
\begin{equation}\label{selfenergy}
 \mathbf{\Sigma}=\left(%
\begin{array}{cccc}
 0 & t_{12} & 0 & t_{13}^0\\
 t_{12}^* & 0 & t_{23} & 0\\
 0 & t_{23}^* & 0 & 0\\
 t_{13}^{0^*} & 0 & 0 & 0
\end{array}\right),
\end{equation}
and 
\begin{equation}
 \mathbf{g}_\s=\left(%
\begin{array}{cccc}
 g_{1\s} & 0 & 0 & 0 \\
 0 & g_{2\s} & 0 & 0 \\
 0 & 0 & g_{3\s} & g_{\underline{3}^*\s} \\
 0 & 0 & g_{\underline{3}\s} & g_{3\s} 
\end{array} \right),
\end{equation}
with the matrix elements\cite{comment4}
\begin{eqnarray}
 g_{j\s}(\tau,\tau') &=& {\sum_{\mathbf{k}_j}}' g_{\mathbf{k}_j\s} (\tau,\tau'), \phantom{x}j=1,2,3\\
 g_{\underline{3}\s} (\tau,\tau') &=& \sum_{\mathbf{k}_3} e^{i \mathbf{k}_3 \cdot \mathbf{R}} g_{\mathbf{k}_3\s} (\tau,\tau'),\\
 g_{\underline{3}^*\s} (\tau,\tau') &=& \sum_{\mathbf{k}_3} e^{-i{k}_3 \cdot \mathbf{R}}  g_{\mathbf{k}_3\s} (\tau,\tau').
\end{eqnarray}

Performing an analytic continuation in Eq. (\ref{dysonmatrix}) we obtain the Dyson equation for the retarded Green function
\begin{equation}\label{matrixGr}
 \mathbf{G}_\s^r = [\mathbf{g}_\s^{r^{-1}}-\mathbf{\S}^r]^{-1},
\end{equation}
and the Keldysh \cite{qf05} equation
\begin{equation}\label{matrixGlesser}
 \mathbf{G}_\s^<=\mathbf{G}_\s^r \mathbf{g}_\s^{r^{-1}} \mathbf{g}_\s^< \mathbf{g}_\s^{a^{-1}} \mathbf{G}_\s^a,
\end{equation}
where
\begin{equation}\label{grlesser}
 \mathbf{g}^{r,<}_\s=\left(%
\begin{array}{cccc}
 g_{1\s}^{r,<} & 0 & 0 & 0 \\
 0 & g_{2\s}^{r,<} & 0 & 0 \\
 0 & 0 & g_{3\s}^{r,<} & g_{\underline{3}^*\s}^{r,<} \\
 0 & 0 & g_{\underline{3}\s}^{r,<} & g_{3\s}^{r,<}
\end{array} \right).
\end{equation}
The advanced Green function $\mathbf{g}^a_\s$ is given by $\mathbf{g}^a_\s=[\mathbf{g}^r_\s]^*$. 
From Eqs. (\ref{matrixGr}) and (\ref{matrixGlesser}) we see that if $\mathbf{g}^r_\s$ and $\mathbf{g}^<_\s$ are known
we can determine immediately $\mathbf{G}^r_\s$ and $\mathbf{G}^<_\s$, and so the spin-resolved current, Eq. (\ref{I1second}). 
The first nonzero (diagonal) elements ($g_{1\s}^<$ and $g_{1\s}^r$) in Eq. \ref{grlesser} are 
\begin{eqnarray}
g_{1\s}^<(\w) &=& \sum_{\mathbf{k}_1} g_{k_1\s}^<(\w)=\sum_{\mathbf{k}_1} 2 \pi i f_1(\w) \delta(\w-\epsilon_{\mathbf{k}_1\s})\nonumber \\ 
          &=& 2 \pi i f_1(\w) \rho_{1\s}(\w),
\end{eqnarray}
\begin{eqnarray}
g_{1\s}^r(\w)&=&\sum_{\mathbf{k}_1} g_{k_1\s}^r(\w)=\sum_{\mathbf{k}_1} \left[P\left(\frac{1}{\w-\epsilon_{\mathbf{k}_1\s}}\right)- i \pi \delta(\w-\epsilon_{\mathbf{k}_1\s})\right] \nonumber \\
         &=& \Lambda_1(\w) - i \pi  \rho_{1\s}(\w),
\end{eqnarray}
where $f_1(\w)$ is the tip Fermi distribution function, $\rho_{1\s}(\w)$ is the tip density of states and $\Lambda_1(\w)=P \sum_{\mathbf{k}_1} \left(\frac{1}{\w-\epsilon_{\mathbf{k}_1\s}}\right)$, where $P$ stands for the Cauchy Principal Value. The retarded adatom Green function $g_{2\s}^r(\w)$ is given by Eq. (\ref{gradatom}). The lesser component can be calculated straightforwardly
from the relation $g_{2\s}^<(\w)=i n_\s A_\s(\w)$,
where
\begin{equation}
 A_\s(\w)=2 \pi (1-n_{\bar{\s}}) \delta(\w-\epsilon_\s) + 2 \pi n_{\bar{\s}} \delta(\w-\epsilon_\s-U),
\end{equation}
and $n_\s$ is the average spin-resolved occupation of the adatom. The third diagonal element of $\mathbf{g}^r$ is given by
\begin{eqnarray}
 g_{3\s}^r(\w) &=& \sum_{\mathbf{k}_3} g_{k_3\s}^r (\w)=\sum_{\mathbf{k}_3} \left[P\left(\frac{1}{\w-\epsilon_{\mathbf{k}_3}}\right)- i \pi \delta(\w-\epsilon_{\mathbf{k}_3})\right] \nonumber \\
         &=& \Lambda_3(\w) - i \pi  \rho_3(\w),
\end{eqnarray}
where $\rho_3(\w)=\rho_3^0$ is the 2D density of states of the surface defined at the end of Sec.~\ref{NRT1}, and
$\Lambda_3(\w)=P \sum_{\mathbf{k}_3} \left(\frac{1}{\w-\epsilon_{\mathbf{k}_3}}\right)$. For the corresponding lesser Green function
we find
\begin{equation}\label{g3lesserfinal}  
 g_{3\s}^<(\w) = \sum_{\mathbf{k}_3} g_{\mathbf{k}_3\s}^< (\w)=2 \pi i f_3(\w) \rho_3^0,
\end{equation}
where $f_3(\w)$ is the Fermi distribution function of the host surface.

Finally, we should calculate the off-diagonal elements of the matrix $\mathbf{g_{\s}}$. For the retarded Green function we have
\begin{eqnarray}
\label{g3r}
 g_{\underline{3}\s}^{r}(\w) &=& \sum_{\mathbf{k}_3} e^{i \mathbf{k}_{3} \cdot \mathbf{R}} g_{k_3\s}^r (\w)\nonumber \\
                                                &=& \rho_3^0 \int_{-1}^1 dx \frac{\frac{\w}{D}-x}{(\frac{\w}{D}-x)^2+\delta^2} J_0(k_F R \sqrt{1+x})-\nonumber\\
                                                 &&\phantom{xxxxxxxxxxx} i \pi \rho_3^0 J_0\left(k_F R \sqrt{1+\frac{\w}{D}}\right).
\end{eqnarray}
For the lesser Green function we find
\begin{eqnarray}
\label{g3<}
 g_{\underline{3}\s}^{<}(\w) &=& \sum_{\mathbf{k}_3} e^{i \mathbf{k}_{3} \cdot \mathbf{R}} g_{k_3\s}^< (\w),\nonumber\\
                                                &=& 2 \pi i \rho_3^0 f_3(\w)  J_0\left(k_F R \sqrt{1+\frac{\w}{D}}\right).
\end{eqnarray}

The Green functions $g_{\underline{3}^{*}\s}^{r}$ and $g_{\underline{3}^{*}\s}^{<}$ have exactly the same expressions as Eqs. (\ref{g3r}) and (\ref{g3<}), respectively.

\subsection{Parameters and a summary of the numerical technique}

Our main task is to determine the current from Eq. (\ref{I1second}). To this end, we first calculate
$\mathbf{G}_\s^r$ and $\mathbf{G}_\s^<$ from Eqs. (\ref{matrixGr}) and (\ref{matrixGlesser}), respectively. Then we substitute
the relevant matrix elements in Eq. (\ref{I1second}). Note that in the presence of the Coulomb
interaction, $g_{2\s}^r(\w)$ and $g_{2\s}^<(\w)$ depend on the adatom occupation $n_{\bar{\s}}$; so do $\mathbf{G}_\s^r$ and $\mathbf{G}_\s^<$.
This implies a self-consistent calculation, where $n_\s$ is calculated iteratively via
\begin{equation}
 n_\s = \int \frac{d\w}{2 \pi i} G^<_{22\s}(\w).
\end{equation}

As a matter of simplification we use the wide-band limit for the tip, so the density of states $\rho_{1\s}(\w)$ is taken as constant $\rho_1^0$ ($\rho_3^0$ is already a constant), evaluated at the Fermi level. This is a good approximation when $eV,k_{B}T<<D$, where $D$ is the band half width.
The ferromagnetism of the tip is introduced via the density of states 
$\rho_{1\s}=\rho_1^0(1 \pm p)$, where $p$ is the tip polarization and the $+$ and $-$ signs apply
to spin up and down, respectively.\cite{fms04} Since the characteristic tunneling rate between the tip and the adatom is given by $\Gamma_{1\s}=2 \pi |t_{12}|^2 \rho_{1\s}$,
we find $\Gamma_{1\s}=2 \pi |t_{12}|^2 \rho_1^0 (1 \pm p)$, which is the standard phenomenology to account for the ferromagnetism
of the electrode.\cite{wr01} Analogously, the tunneling rate between the adatom and the host surface is
$\Gamma_3=2 \pi |t_{23}|^2 \rho_3^0 \equiv \Gamma_0$. In our calculation we take $\Gamma_0=10 \mu$eV as the energy scale.
All the phenomenological parameters used in this paper are summarized in table \ref{tabela1}.
\begin{table}[htb]
\begin{center}
\begin{tabular}{cc}
 \hline
 \textbf{Parameter} \  & \qquad \textbf{Magnitude}\\
 \hline \hline
 {\small Band half width}\hfill\hfill\hfill &\qquad $D=1000\Gamma_0$ \hfill\hfill  \\
 {\small Adatom Charging Energy}\hfill\hfill\hfill &\qquad $U=30\Gamma_0$ \hfill\hfill \\
 {\small Tip chemical potential}\hfill\hfill\hfill & \qquad $\mu_1=-eV/2$ \hfill\hfill \\
 {\small Host chemical potential}\hfill\hfill\hfill & \qquad  $\mu_3=eV/2$ \hfill\hfill \\
 {\small System temperature}\hfill\hfill & \qquad $k_B T=\Gamma_0$\hfill\hfill \\
 {\small Tip degree of polarization}\hfill\hfill &\qquad  $p=0.4$ \hfill\hfill \\
 {\small Decaying factor of $t_{12}$}\hfill\hfill &\qquad  $R_0=1/k_F$ \hfill\hfill\\ \hline \hline
\end{tabular}
\caption{Parameters used in the self-consistent calculation.}
\label{tabela1}
\end{center}
\end{table}

Note that in Table \ref{tabela1} we define the origin of the energy scale such that $\mu_1=\mu_3=\epsilon_{\sigma}=0$ at zero bias, i.e., the adatom energy level is aligned with the chemical potential of the leads in the absence of an applied bias.
In order to be consistent with $\Gamma_0 = 2 \pi |t_{23}|^2 \rho_3^0$ and the value of $\rho_3^0$ for typical semiconductors, e.g, GaAs, the factor $t_{23}$ is taken as $\Gamma_0/\sqrt{10}$. We assume $t_{13}^0=0.01t_{23}^0$ and adopt values for $t_{12}^0$ and $\rho_1^0$ consistent with $\Gamma_0=2\pi|t_{12}^0|^2\rho_1^0$.
Note that for $eV=\pm 30 \Gamma_0$ the adatom can be occupied by a single electron since $\epsilon_\s$ is within the conduction window
(the energy range between $\mu_1$ and $\mu_3$) and $\epsilon_\s+U$ is without this range. On the other
hand for $eV=\pm 150 \Gamma_0$ the dot can be doubly occupied since both $\epsilon_\s$ and $\epsilon_\s+U$ lie inside the conduction window.
The parameter $R_0$ controls how fast the coupling $t_{12}$ decays in space when the tip moves away from the adatom. We take
it equal to $k_F^{-1}$. Hereafter $k_F^{-1}$ will be used as a length scale.

\section{Results}

\subsection{Single Occupancy}

\begin{figure}[b]
\par
\begin{center}
\epsfig{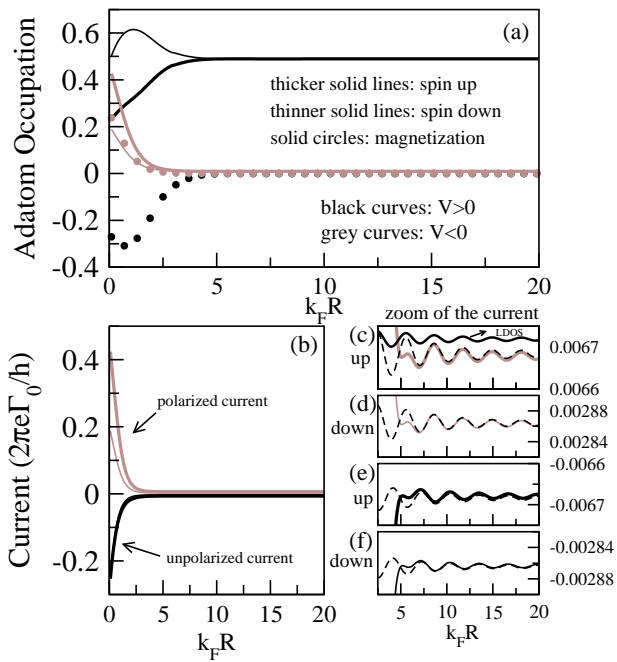}
\end{center}
\caption{(a) Adatom occupations and magnetization and (b) spin-resolved currents against the tip position $R$.
Both negative (grey line) and positive (black line) bias voltages are shown. For $|eV|=30\Gamma_0$
only single occupancy (i.e., $\epsilon_\s+U$ is much higher than the emitter chemical potential) is allowed. 
This results in the spin-diode effect: for negative bias voltages
the current is polarized for all $R$ values while for positive biases the current polarization goes to zero
when the tip is close to the adatom. Inset: zoom of each spin component of the current in 
the range where the tip-adatom coupling is negligible. In (c) we also show the LDOS (thick solid black line)
evaluated at $\epsilon_F$.
The LDOS displays Friedel oscillations which show up in the polarized current. The vertical scale for the LDOS
is not shown.} 
\label{fig2} 
\end{figure}

Figure \ref{fig2}(a) shows the adatom occupations against the lateral distance $R$ between the tip and the adatom. We study both forward
($eV=+ 30\Gamma_0$) and reverse ($eV=- 30\Gamma_0$) bias cases. As mentioned before, for $|eV|=30\Gamma_0$ the adatom cannot be doubly occupied, since $\epsilon_{\s}+U$ lies above the Fermi energy of the source. For $V>0$ the host NM surface is the source and the FM tip is the drain of electrons, i.e., the electrons flow from the NM surface ($+$ adatom) to the tip. For $V<0$ we have the opposite. 

For $R=0$ and $V>0$ (black lines) we find $n_\uparrow < n_\downarrow$. This is reasonable since spin up electrons tunnel from the adatom into the tip much more easily than the spin down ones, due to the larger spin up density of states, $\rho_{1\s}=\r_1^0(1 \pm p)$ ($p>0$), which implies $\Gamma_{1\uparrow}>\Gamma_{1\downarrow}$, and gives rise to a larger spin down population. For a constant bias voltage, as the tip moves away from the adatom [see Eq. (\ref{t12})] the tunneling rates $\Gamma_{1\uparrow}$ and $\Gamma_{1\downarrow}$ decrease, but the incoming rates $\Gamma_{3}$ stay the same, this results in an increase of both $n_\uparrow$ and $n_\downarrow$. We note, however, that the spin down population (thinner solid black line) increases more quickly than the spin up one (thicker solid black line), this is because $\Gamma_{1\downarrow}<\Gamma_{1\uparrow}$, see Fig. \ref{fig2}(a). As the $n_{\uparrow}$ adatom population increases with $R$, the $n_{\downarrow}$ one tends to be more blocked due to the spin-dependent Coulomb blockade. The interplay between the Coulomb blockade and the decrease of the tunneling rates $\Gamma_{1\uparrow,\downarrow}$ makes $n_{\downarrow}$ reach a maximum, subsequently decreasing to attain the limit $n_\uparrow=n_\downarrow=0.5$ for large enough $R$'s.

In contrast, for $eV=-30 \Gamma_0$, $n_\uparrow > n_\downarrow$ for small $R$ values, see in Fig. \ref{fig2}(a) the solid thicker and thinner grey lines. 
This is a consequence of $\Gamma_{1\uparrow}>\Gamma_{1\downarrow}$, which means that more spins up tunnel to the adatom. Besides, 
the outgoing rates $\Gamma_3$ (the same for the up and down components) is smaller than $\Gamma_{1\uparrow}$, which results in a larger 
spin up accumulation in the adatom. As the tip moves away from the adatom, $\Gamma_{1\uparrow}$ and $\Gamma_{1\downarrow}$ go to zero 
exponentially and the populations $n_\uparrow$ and $n_\downarrow$ are completely drained out into the host surface,
thus resulting in an empty adatom.

The magnetization $m=n_\uparrow-n_\downarrow$ is also shown in Fig. \ref{fig2}(a) (solid circles). Observe that for small $R$ the adatom is
spin down polarized for $V>0$ and spin up polarized for $V<0$. As $R$ increases, $m$ tends to zero for both positive and
negative bias voltages. However, $m$ tends to zero much slower for $V>0$ than for $V<0$, a consequence of
the interplay between the Coulomb interaction (spin-Coulomb blockade) and the tunneling rates $\Gamma_{1\s}$, that change with the tip position as it moves away from the adatom.  

In Fig. \ref{fig2}(b) we present the spin-resolved currents for both $eV=\pm30\Gamma_0$. The spin-diode effect\cite{fms07} can be clearly
seen for small values of $R$. While for $V>0$ (black lines) we find $I_\uparrow \approx I_\downarrow$ for small
$R$ values, for $V<0$ (grey lines) we observe $I_\uparrow > I_\downarrow$. This shows that the current polarization can be controlled
via both the bias sign and the tip position. In the case of $V<0$, we have $\Gamma_{1\uparrow}>\Gamma_3>\Gamma_{1\downarrow}$, i.e., the spin up population is greater than the spin down one, $m>0$. As a consequence, in the absence of the Coulomb interaction in the adatom, $I_{\uparrow}>I_{\downarrow}$ (the case $U=0$ resembles the curves in the double occupancy regime ($eV>>U$), see Fig. \ref{fig3}. In the presence of $U$, $I_{\downarrow}$ is suppressed, since $n_{\downarrow}$ tends to be more blocked than $n_{\uparrow}$ [see Fig. \ref{fig2}(a)], which results in an enhancement in the difference between $I_{\uparrow}$ and $I_{\downarrow}$. For $V>0$, the magnetization changes sign $m<0$, now the spin up population tends to be more blocked, and $I_{\uparrow}$ is more strongly suppressed compared to $I_{\downarrow}$, interestingly attaining values close to $I_{\downarrow}$. The amplification of $I_\uparrow$ compared to $I_\downarrow$ for $V<0$, when the
tip is closer to the adatom, does not occur in the double occupancy regime ($eV= \pm 150 \Gamma_0$) as we will see in the next section.

In Fig. \ref{fig2}(c)-(f) we show the current for a range of $R$ in which only the direct tip-host
tunneling (non-resonant transport) is relevant. Note that $I_\uparrow$ and $I_\downarrow$ tend to distinct plateaus for large enough $R$'s. 
These plateaus correspond to the background current between the tip and the host surface, given approximately by Eq. (\ref{Ianalytic}).  
By comparison with Eq. (\ref{I1second}) we plot in dashed line the current obtained 
via Eq. (\ref{I1T}). In the large-$R$ limit, we expect an agreement between both equations, since Eq. (\ref{I1T}) was derived in the case of negligible tunneling between the tip and the adatom (see the solid black and grey lines). The minor difference between the two results is due to Eq. (\ref{I1T}) having been obtained in the limit of small tip-surface coupling parameter $t_{13}^0$. \cite{comment2} The LDOS evaluated at the Fermi level, $\r_{\s}(R,0)$, is also shown in Fig. \ref{fig2}(c); it oscillates around the unperturbed surface density of states $\rho_3^0$. Friedel-like oscillations are seen for both spin components, thus reflecting the oscillations in
the LDOS due to the scattering center (adatom). Note that Friedel oscillations have been seen experimentally in a variety of systems.\cite{pa94,kk01,ia08,sk10}
 
\subsection{Double Occupancy}

\begin{figure}[h!]
\par
\begin{center}
\epsfig{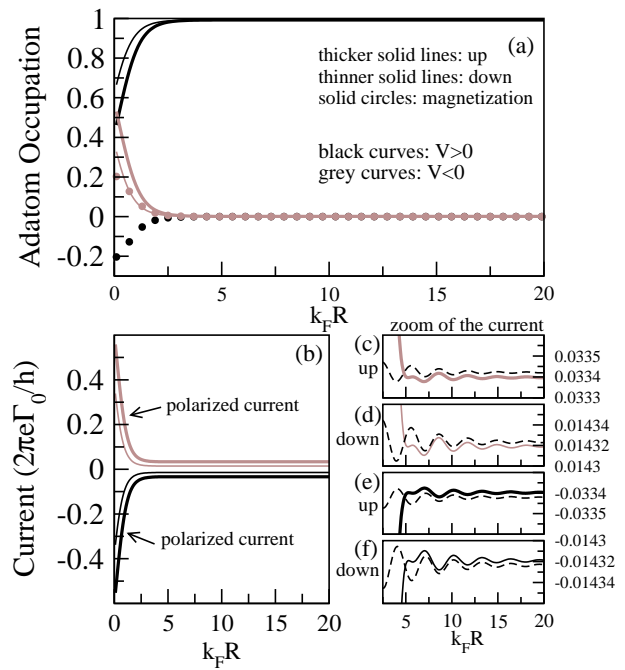}
\end{center}
\caption{Similar to Fig. \ref{fig2} except that $|eV|=150 \Gamma_0$ here. This bias allows double occupancy of the adatom. 
In this regime no spin-diode effect is observed. The spin polarized currents are the same
(in modulus) for both positive and negative $eV$. For forward voltages (\textit{black} curves)
the adatom becomes doubly occupied for large $R$ values ($n_\uparrow+n_\downarrow=2$).} \label{fig3}
\end{figure}

\vspace{0.6cm}

Figure \ref{fig3} shows the spin-resolved (a) adatom occupations and (b) currents in the double
occupancy regime, i.e., when the bias voltage is large enough ($|eV|=150 \Gamma_0$) to allow for two electrons of opposite spins
in the adatom at the same time. For $V>0$ (black lines), as $R$ increases both $n_\uparrow$ and $n_\downarrow$
populations tend to unity and the adatom becomes doubly occupied ($n_\uparrow+n_\downarrow=2$). This is so
because electrons can jump into the adatom but cannot leave it for large $R$ values. In contrast,
for $V<0$ both $n_\uparrow$ and $n_\downarrow$ vanish as $R$ increases because the electron source
(tip) decouples from the adatom. Observe also that the magnetization $m$ is enhanced as $R$ tends to zero and switches sign depending on whether $eV$ is positive or negative.

The current in the double occupancy regime [Fig. \ref{fig3}(b)] has a similar behavior for
both positive and negative biases. Note that $|I_\uparrow|>|I_\downarrow|$ for $V>0$ and $V<0$, in contrast
to the single occupancy regime where we find $I_\uparrow \approx I_\downarrow$ for $V>0$ [Fig. \ref{fig2}(b)]; hence no spin-diode effect is observed here. In Figs. \ref{fig3}(c)-(f) we show a 
zoom of the spin-resolved currents in the range of negligible tip-adatom coupling. As before, we observe
Friedel oscillations, which reflect the disturbance in the LDOS due to the localized impurity (adatom). 
The dashed black lines in Figs. \ref{fig3}(c)-(f) show the current obtained via Eq. (\ref{I1T}).

\subsection{Current Polarization}

\begin{figure}[h!]
\par
\begin{center}
\epsfig{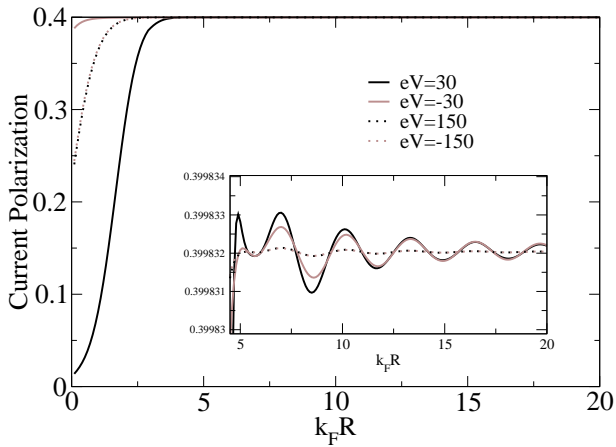}
\end{center}
\caption{Current polarization against $R$. In the single occupancy regime ($eV=\pm 30 \Gamma_0$) the
current polarization is approximately constant for reverse bias ($V<0$) and goes down to zero
for direct bias ($V>0$) when the tip is close to the adatom. In the case of double
occupancy ($eV=\pm 150 \Gamma_0$) the current polarization is suppressed for small $R$, however it
does not vanish; notice that the dotted lines lie essentially on top of each other. For large $R$'s, the polarization for both single and double occupancies tends to a plateau and presents tiny oscillations (inset).} \label{fig4}
\end{figure}

Figure \ref{fig4} displays the current polarization, $\wp=(I_1^\uparrow-I_1^\downarrow)/(I_1^\uparrow+I_1^\downarrow)$, as a
function of $R$. Both single ($|eV|=30 \Gamma_0$) and double ($|eV|=150 \Gamma_0$) occupancies are shown. For $eV=-30 \Gamma_0$  
(solid grey line) the polarization is approximately constant $\sim 40\%$ while for $eV=+30 \Gamma_0$
(solid black line) it is strongly suppressed when the tip is close to the adatom. 
This is a type of \textit{spatially resolved} spin-diode\cite{fms07}, which allows a polarized current to flow only for reverse bias.
In the case of double occupancy, though, both positive and negative biases present a similar behavior with a $40\%$ current polarization away from the adatom and a slight suppression as the tip moves closer to it. This $40\%$ current polarization for large $R$'s in the single- \textit{and} double-occupancy cases follows straightforwardly by calculating $\wp$ using the spin-resolved non-resonant currents in Eq. (\ref{Ianalytic}). The inset shows a blow up of the current polarization
and also reveals Friedel oscillations.
\vspace{0.5cm}

\section{Conclusion}

We have studied spin-polarized quantum transport in a system composed of a FM STM tip coupled
to a NM host surface with a single adsorbed atom. Due to Coulomb interaction in the adatom the
system can operate as a spin-diode when the tip is nearby the adatom. In the singly occupied case
and direct bias ($V>0$), the current polarization can vary from zero up to 40\% depending on the tip position.
For reverse bias, though, the polarization is pinned close to 40\% for all tip positions. In the double
occupancy regime the current polarization is the same for both forward and reverse biases,
with a slight suppression as the tip moves closer to the adatom. Additionally, the adatom magnetization
can be tuned by varying the tip position and its sign can switch depending on the bias. 
Finally, we have also found spin-resolved Friedel oscillations in the current as the tip moves laterally
away from the adatom, thus reflecting the oscillations in the surface LDOS induced by the adatom acting like an effective impurity.

\section*{Acknowledgments}

Two of the authors JCE and PHP acknowledge useful discussions with D. Loss.
The authors acknowledge financial support from CNPq, CAPES, FAPEMIG and FAPESP.


\begin{thebibliography}{99}
\bibitem{comment1} The proportionality $t_{13} \propto e^{i\mathbf{k_3}\cdot\mathbf{R}}$ was originally discussed
in Ref. \onlinecite{jt85} and further used in Ref. \onlinecite{mp01}.
\bibitem{comment2} The current plateau given by Eq. (\ref{Ianalytic}) is valid only for small tip-surface
coupling parameter $t_{13}^0$. If we calculate the Green functions $G_{\s}^r(\mathbf{R},\mathbf{R},\w)$ and $G_{\s}^<(\mathbf{R},\mathbf{R},\w)$
accounting for the tip, a small correction arises and Eq. (\ref{Ianalytic}) becomes $I_1^\s=G_\s V$, where $G_\s=\frac{e}{h}\frac{4r}{(1+r)^2}$, with $r=\pi^2 |t_{13}^0|^2 \rho_{1\s} \rho_3$. This feature was pointed out in Ref. \onlinecite{ev09}. Note that we can regain Eq. (\ref{Ianalytic}) (now divided by $\hbar$) by expanding the above expression for the current up to second order in the coupling parameter.
\bibitem{comment3} To obtain the equation of motion for $G_{\mathbf{k}_3 \mathbf{k}'_3,\s}^r(t-t')$, we first perform a time derivative with respect to $t$; the resulting equation couples $G_{\mathbf{k}_3 \mathbf{k}'_3,\s}^r(t-t')$ to $G_{\mathbf{k}_2 \mathbf{k}'_3,\s}^r(t-t')$. By deriving this second Green function now with respect to $t'$, we find an equation that relates $G_{\mathbf{k}_3 \mathbf{k}'_3,\s}^r(t-t')$ to the adatom Green function $G^r_{\mathbf{k}_2 \mathbf{k}'_2,\s}(t-t')$.
\bibitem{comment4} Note that Eq. (27) and the LDOS, given by Eq. (32), can be derived from the present matrix formulation by taking $t_{12}=0$ and $t_{13}=0$ in the self-energy matrix [Eq. (\ref{selfenergy})].
\end{thebibliography}
\end{document}